\documentstyle[preprint,aps]{revtex}
\draft
\begin{document}
\title{Nonuniversal Critical Spreading in Two Dimensions}
\author{Ronald Dickman$^{\dagger}$ }
\address{Department of Physics and Astronomy, Lehman College, CUNY,
Bronx, NY 10468-1589 }
\date{\today}
\maketitle
\begin{abstract}
Continuous phase transitions are studied in a two dimensional
nonequilibrium model with
an infinite number of absorbing configurations.  Spreading from a
localized source is characterized by
{\em nonuniversal} critical exponents, which vary continuously
with the density $\phi $ in the surrounding region.
The exponent $\delta$ changes by more than an order of magnitude,
and $\eta$ changes sign.  The location of the
critical point also depends on $\phi$, which has important implications
for scaling. As expected on the basis of
universality, the static critical
behavior belongs to the directed percolation class.
\vspace {0.3truecm}

{PACS numbers: 05.50.+q, 02.50.-r, 05.70.Ln
}
\end{abstract}
\vspace{1.0truecm}

\noindent $^{\dagger}${\small e-mail address:
dickman@lcvax.lehman.cuny.edu }\\
\vspace{1em}

\section{Introduction}

Despite recent progress, understanding
the phase diagrams and critical points of
nonequilibrium models remains a challenging problem.
In the relatively well-understood case of
continuous transitions into an absorbing state
(i.e., one that traps the dynamics),
a high degree of
universality has been found, supporting the prediction
that such transitions belong generically to the class of directed
percolation (DP) \cite{janssen1,pgrg,geoff}.
Examples include the basic contact process and its variants
\cite{harris,liggett,durrett,konno,tdpdif,tdpcpgen},
surface reaction models \cite{geoff,zgb,iwanzgb},
branching annihilating random walks with
odd parity \cite{taktre,iwanbaw,baw4ann}, and
assorted multiparticle processes \cite{aukrust,BBC,tripann,tripcrea,park}.
The {\em static} critical behavior of models with {\em multiple}
absorbing configurations
\cite{iwanpcp,pcppre,albano,ditri,yald,iwancono,iwandt,mendes}
also falls in the DP class.
On the other hand, some aspects of {\em time-dependent}
critical behavior in the latter class of models are surprising:
the critical exponents which describe spreading from a seed
vary continuously with the density in the environment, and the
usual hyperscaling relation must be modified.
Until now this phenomenon has been investigated in one dimension only;
here I present the first study of nonuniversal critical spreading in a
two dimensional model.

Models with an absorbing state are characterized by an
order parameter $\rho$ --- the density of self-reproducing entities
or {\em active sites}.  The stationary value $\overline{\rho}$
vanishes as the reproduction rate $\lambda$ approaches a
critical value $\lambda_c$: $\overline{\rho} \propto \Delta^{\beta}$
for small $\Delta = \lambda - \lambda_c $.
For $\lambda < \lambda_c$ the only steady state is the absorbing
state, in which all change ceases.
Much insight is gained from
studies of spreading from a single active site \cite{torre}.
The chief quantities of interest are the survival probability, $P(t)$,
(the process is said to survive so long as it has not
become trapped in the absorbing state), the
mean number of active sites $n(t)$, and the mean-square distance,
$R^2(t)$, of active sites from
the original seed.  At the critical point they follow asymptotic
power laws,

\begin{equation}
\label{survpwr}
P(t) \propto t^{-\delta},
\end{equation}
\begin{equation}
\label{poppwr}
n(t) \propto t^{\eta},
\end{equation}
and
\begin{equation}
\label{sprpwr}
R^{2}(t) \propto t^{z}.
\end{equation}

\noindent These power laws, which form the most salient feature
of the critical spreading process,  have been confirmed
for a broad class of models.
The exponents $\delta$, $\eta$, and $z$,
estimated from simulations and series expansions
\cite{tdpjsp} are in good agreement with DP values.

Several models have been devised with
absorbing states that embrace a large number of configurations:
two-dimensional catalysis models \cite{albano,ditri,yald,iwancono},
and one-dimensional particle systems
such as the pair contact process (PCP) \cite{iwanpcp,pcppre,mendes}.
In the PCP each site of a lattice ({\bf Z}) is either
vacant or occupied by a particle;
the active ``sites" are nearest-neighbor particle pairs.
They are annihilated with probability $p$,
and give birth, with probability $1-p$, to a new particle at a
vacant neighboring site. Since particles may only be
created or destroyed if there are pairs,
and since particles cannot move to
other sites, any configuration of isolated particles
is absorbing; the system always becomes trapped in such a
configuration if $p > p_c$.
While static critical behavior in models with multiple
absorbing configurations still conforms
to DP \cite{munoz}, spreading presents a
new feature: one can choose the absorbing
configuration into which the seed is placed.
Critical spreading has been studied in three
rather different one-dimensional
models with multiple absorbing configurations: the PCP, the
threshold transfer process, and the dimer reaction \cite{pcppre,mendes}.
For ``natural'' absorbing configurations (those generated by
the critical process itself), the spreading exponents
assume DP values.  If, however, the
initial density $\phi_{\rm i}$ differs
from the natural value, the exponents $\delta$ and $\eta$ change;
they vary linearly with $\phi_{\rm i}$.
($\phi$  represents the density of
isolated particles in the PCP, {\em not} the order parameter.)
A new exponent, $\beta'$ must be introduced to
describe the ultimate survival probability:

\begin{equation}
P_{\infty} \equiv \lim_{t \rightarrow \infty}
P(t) \propto (p_{\rm c} - p)^{\beta'}
\label{betaprime}
\end{equation}
in the active regime.  This exponent
also depends on $\phi_{\rm i}$. ($\beta' = \beta$ for natural initial
configurations, just as for models with a unique absorbing
configuration.)  It
appears in the generalized hyperscaling relation \cite{mendes},

\begin{equation}
\label{ghyper}
(1 + \frac{\beta}{\beta'}) \delta +  \eta = \frac{dz}{2},
\end{equation}

\noindent which reduces to the relation originally derived by
Grassberger and de la Torre \cite{torre} for $\beta' = \beta$.
Certain properties are {\em independent}
of $\phi_{\rm i}$: the location of
the critical point, the exponent $z$, and $\delta + \eta$, which
governs population growth in surviving trials.  This suggests
that the initial density influences the survival probability,
reflected in the exponents $\delta$ and $\beta'$, but does not
alter the asymptotic properties of surviving trials.

The background sketched above motivates the investigation of
critical spreading
in dimensions $\geq 2$,  which offers a greater range of possible growth
patterns.  In the following section
I define a two-dimensional model with multiple absorbing configurations,
and describe the simulation procedure used to
study it.  Results are presented in
Sec. III, and a scaling analysis in Sec. IV, followed by a brief summary in
Sec. V.

\section{The Model}

The second neighbor reaction (SNR) is a Markov
process or {\em interacting particle system}
in which each site of the square lattice {\bf Z}$^2$ is
either vacant or occupied by a particle.
Particles may not occupy adjacent sites. If all four nearest
neighbors of a vacant site (i,j) are also vacant (i,j) is said to be
{\em open}.  Open sites are the only active sites in the model.
In each step of the process, a site is chosen at random;
if it is open, a particle is placed there provisionally, otherwise nothing
happens.  Suppose a particle has just arrived at (i,j).
If all four of the second neighbors (i$\pm $1, j$\pm $1)
are vacant the new particle is adsorbed, and remains at (i,j).
But if any of the second neighbors are occupied,
the new particle is only adsorbed with probability $p$;
with probability $1-p$ it reacts
with one of its neighbors
(selected at random, if necessary), and both
particles leave the lattice.  Each such reaction creates one or more new
open sites. Adsorbed particles are immobile, and can only
leave the surface in a reaction with a newly-arrived particle.
Each adsorption
attempt marks a fixed time interval; we take the time unit as $N=L^2$
such events.

Since the fraction of open sites is quite small
in the vicinity of the critical point, efficiency is greatly improved
by choosing trial adsorption sites
from a list of open sites.  This entails a variable time increment,
$\Delta t = 1/N_o$, for
each event, where $N_o$ is the number of open sites.  Taking statistics
at fixed time intervals (rather than after a fixed number of events),
ensures proper stationary averages.

For $p=1$ the model corresponds to random sequential
adsorption (RSA) with nearest-neighbor exclusion; starting
from an empty lattice, particles
adsorb until no open sites remain.
In this case there is an exponential
approach to the jamming density $\phi_{\infty} =
0.36413$ \cite{rsamc,rsaser}.
The order parameter $\rho$ for the SNR is the fraction of open sites;
configurations devoid of open sites are absorbing.
Of course, many absorbing configurations are possible, with
particle densities $\phi$ ranging from 1/5 to 1/2 (see Fig. 5).
I shall refer to the set of absorbing configurations characterized by
common statistical properties (particle density and correlations) as an
{\em absorbing state}.  It is also useful to define the ensemble
${\cal A} _{p,L}$ of absorbing
configurations generated by the SNR with adsorption
probability p, starting from an empty lattice of size $L$.
For any absorbing configuration $\zeta$ on $L^2$ sites,
${\cal A} _{p,L}(\zeta)$ is the probability of realizing $\zeta$ as
the final configuration.

Based upon experience with models of this kind --- in
particular, with the dimer reaction, the one-dimensional
cousin of the SNR \cite{pcppre,mendes} --- one expects that
there is a critical value $p_c$ above
which the system always gets trapped, whilst for $p<p_c$ there is an
active state in which
the reaction may proceed indefinitely.  Equivalently, one expects that
for $p>p_c$ the absorbing state is the only stationary state, and that
for $p<p_c$ there is in addition a nontrivial invariant
measure on configuration space.  (While the only latter emerges
in the infinite-size limit, for $p < p_c$ the
quasi-stationary state is sufficiently
long-lived to allow precise characterization in simulations.)

\section{Simulation Results}

An accurate estimate of $p_c$ is crucial for reliable determination
of critical behavior.
Several simulation methods are available for locating the
critical point: half-life studies, time-dependent
simulations, and analysis of stationary properties.
Use of any one method exclusively
can be misleading.  Here I begin with the
half-life method,  which is based on finite-size scaling analysis of
the mean survival time. One generates a large number of independent
realizations, all starting from an empty
lattice (all sites open), with periodic boundaries,
and determines the time $\tau$ required for half
the sample to reach the absorbing state.
$\tau$ is a decreasing function of $p$
and an increasing function of lattice size $L$.
One expects that
in the vicinity of the critical point \cite{aukrust}

\begin{equation}
\tau(p,L) \simeq L^{\nu_{||}/\nu_{\perp}} h(\Delta L^{1/\nu_{\perp}}) \; ,
\end{equation}
where $\Delta = p_c - p $ and
$h$ is a scaling function.
The exponents $\nu_{||}$ and $\nu_{\perp}$ govern the divergence of the
relaxation time $t_r$ and the correlation length $\xi$ as
$\Delta \rightarrow 0$: $t_r \propto \Delta^{-\nu_{||}}$ and
$\xi \propto \Delta^{-\nu_{\perp}}$.
The critical point is distinguished by the
simple power-law $\tau \propto L^{\nu_{||}/\nu_{\perp}} $
as $L \rightarrow \infty $.
Following a preliminary survey which indicated that $p_c \simeq 0.3935$,
I estimated $\tau$ for $L$ = 16, 32, 48 64, 96, 128, and 176, for $p$
= 0.3930 --- 0.3940.  It is convenient to plot
$\ln \tau - (\nu_{||}/\nu_{\perp}) \ln L $ versus $\ln L$, using
the expected value of the exponent ratio for DP
in 2+1 dimensions (see Fig. 1).
Only for $p$ in the range  0.3926-0.3927
are the data consistent with a power law.  The corresponding
exponent is $\nu_{||}/\nu_{\perp} = 1.76(2)$, in good agreement
with the DP value, 1.764(26).

More precise estimates of $p_c$ may be obtained from simulations
of spreading \cite{torre}.  As noted above, the
critical point is marked by asymptotic
power laws, Eqs. (\ref{survpwr}) - (\ref{sprpwr}).
This part of the study focuses on the natural absorbing
state --- configurations drawn from ${\cal A} _{p_c,L}$, defined
in Sec. II. Initial configurations are generated by
running the simulation (starting with all sites vacant), at a given $p$,
on a lattice of size $L' = 128$, until it reaches an absorbing
configuration. Following the procedure of Ref. \cite{pcppre},
copies of this configuration are
assembled into an absorbing configuration of size $L=1280$.
Then a single (randomly chosen) particle is removed, creating
an open site (or possibly a small cluster of open sites), which
serves as the seed for the spreading process.

In spreading simulations the process halts
before reaching the boundary; there are no finite-size effects as such,
though one does of course expect finite-time corrections to scaling.
Trials ran to a maximum time $t_{max} = 10^4$;
 $5 \times 10^4$ independent trials were generated for
each $p$ value studied. The survival probability,
mean number of open sites, and mean-square spread for $p=0.392615$
are plotted in Fig. 2, showing
clear evidence for power laws.  In order to
fix $p_c$ and the exponents with precision, I plot, in Fig. 3, the local
slopes of the log-log graphs {\em versus} $1/t$,
for several values of $p$.  (The local slope
$\delta_t$ is found from a linear least-squares fit to the
data for $\ln P(t)$ over the range [$\ln t - 1.5,\; \ln t + 1.5$],
and similarly for
the other exponents.)  I conclude that $0.39261 < p_c < 0.39263$, given
the marked curvatures of the $\delta_t$ and $\eta_t$ graphs
for $p$-values outside this range.
The exponents obtained by extrapolating the local slopes
to infinite-$t$ are in good agreement with accepted values for DP
in 2+1 dimensions (see Table I).

I turn to the exponent $\beta'$ governing the ultimate survival
probability starting from a localized active region.
Natural absorbing configurations, constructed as above (with $p = p_c
= 0.39262$), were used in
determinations of $P_{\infty} $ for
$p<p_c$.  These studies yield $\beta' = 0.58(2)$, whilst in
DP $\beta = \beta' = 0.586(14) $ \cite{brower}.
The stationary open site density $\overline{\rho}$ was determined
on lattices of size $L = 16, 32, 64$, and 128.  To minimize finite size
effects,  $\overline{\rho}(p) $ was obtained for two or more
system sizes, until doubling $L$ produced no significant change.
A linear least-squares fit to a plot of $\ln \rho $ {\em versus}
$\ln (p_c - p)$, (with $p_c = 0.39262$), yields
$\beta = 0.56(2)$.  In summary, the static critical behavior and
critical spreading from ``natural" absorbing
configurations are both fully consistent with DP scaling
in 2+1 dimensions.

It is of interest to characterize the natural absorbing configurations.
Large samples of the latter were generated by running the
simulation at $p_c$, starting from an open lattice ($L \leq 176$),
until it reached an absorbing configuration.
The results for the particle density $\phi_{nat}$ show only a weak
dependence on $L$, permitting reliable extrapolation
to infinite-$L$, yielding $\phi_{nat} = 0.33429(3)$.
Further insight into absorbing configurations is provided by
the two-point total correlation function,

\begin{equation}
h({\bf r}) \equiv \frac{\langle \sigma({\bf x}) \sigma({\bf x + r})
\rangle}{ \phi ^2} - 1,
\end{equation}
where $\sigma ({\bf x})$ is an indicator variable which is 1 iff
site ${\bf x}$ is occupied.  Fig. 4 shows that
$h(r)$ is negligible for $r > 6$.  Thus correlations in natural
absorbing configurations in the SNR are of only slightly greater range
than in the one-dimensional dimer reaction \cite{pcppre}.

Having established the properties of the
active stationary state, the natural
absorbing state, and of critical spreading from the latter, I turn to
spreading in ``non-natural" absorbing configurations.
Detailed studies were performed for the maximal
density ($\phi = 0.5$),
the minimal one ($\phi = 0.2$), and for two kinds of
configurations with $\phi = 0.25$, one
with square-lattice symmetry, the other having a
preferred axis.  In addition to
these periodic states, absorbing configurations were generated
via RSA.
(Examples are shown in Fig 5.)
In each case, the critical
point, $p_c(\phi)$, was identified by searching for asymptotic
power-laws as in Eqs (\ref{survpwr}) ---
(\ref{sprpwr}).

Consider first the lowest density absorbing configuration ($\phi = 0.2$),
in which particles are arrayed in a $\sqrt 5 \times \sqrt 5 $ pattern.
Simulations indicate
that the spreading process is {\em supercritical} for
$p =p_{c, bulk} = 0.39262$. The survival probability
$P(t)$, for example, tends to a nonzero value as $t \rightarrow \infty$,
rather than exhibiting power-law decay.
Power-law spreading occurs for $p = 0.4102(2)$,
about 4\% greater than $p_{c, bulk}$
(see Fig 6).  As indicated in Table I,
the spreading exponents $\eta$ and
$z$ are much larger than the corresponding DP values, and $\delta $
and $\beta'$ are much smaller.
The changes in $p_c$ and in the exponents accord
with the notion that open sites spread more readily in a region of
low particle density.

\begin{table}
 \begin{center}
  \caption{Critical point and exponents in the second
neighbor reaction.  Figures in
parentheses denote uncertainties.}
  \label{tab1}
  \begin{tabular}{|l|l|l|l|l|l|}
$ \phi $ &  $p_c $ & $\delta  $ & $\eta$ & $ z $ & $\beta'$  \\ \hline
0.2         & 0.4102(2)  & 0.078(8)  & 0.64(4) & 1.84(2) &  0.125(5) \\
0.25 (isotropic) & 0.3965(1)  & 0.054(2)  & 0.90(1) & 1.76(1)
&  0.18(5)  \\
0.25 (aniso.) & 0.39238(2) & 0.677(5)  & -0.035(5) &
0.97(1)  &              \\
0.33443 (natural)  & 0.39262(1) & 0.468(6)  & 0.216(8)  &
1.117(13)  &  0.58(1) \\
0.36413 (RSA)  & 0.39240(5)  & 0.64(1)  & 0.035(20) & 1.038(3) &
0.70(2)  \\
0.5          & 0.39220(5) & 1.30(5)  & -0.12(6) & 0.91(2)  &
2.0(2)         \\ \hline
 DP         &    ---           & 0.460(6) & 0.214(8) & 1.134(4) &
0.586(14) \\
  \end{tabular}
 \end{center}
\end{table}

The absorbing configuration with maximal
particle density ($\phi = 1/2$),
by contrast, presents hostile terrain
for spreading.  Running at the
bulk critical point, the process appears to be {\em subcritical}:
$P(t)$ and $n(t)$ decay exponentially.  To
observe power-law spreading one
must reduce $p$ to 0.3922, about 0.14\% below
the bulk critical value.
The shift in $p_c$ is small,  but the effect on the exponents
is dramatic: spreading is sub-diffusive ($z < 1$),
and $\eta $ is negative! (The latter is consistent
with known constraints: $\eta + \delta \geq 0$, so the
population in {\em surviving} trials increases with time.)
Studies with intermediate initial densities,
summarized in Table I and Fig. 7,
confirm the tendencies noted for the extreme cases: higher densities
correspond to smaller values of $\eta$ and $z$,
larger $\delta$ and $\beta'$, and a depressed $p_c$, and conversely,
consistent with the expectation that the larger $\phi$, the more
spreading is impeded.  The only exception is
the anisotropic system, which despite having $\phi < \phi_{nat}$,
resembles isotropic systems with $\phi > \phi_{nat}$.
Why anisotropy renders spreading more difficult is unclear.
The contrast between the two studies at $\phi = 0.25$ is nonetheless
striking, and demonstrates that factors other than the particle
density can influence critical spreading.

\section{Scaling Theory}

We have seen that as $\phi$, the density of particles in the
environment, is varied, the location of the critical point
changes as well. The shift in $p_c$ implies that
critical processes starting from non-natural absorbing configurations
do not evolve to the bulk critical state, which exists
uniquely at $p_{c,bulk}$.
As a result, the usual scaling theory for absorbing states must be
modified.
In (isotropic) absorbing configurations with $\phi < \phi_{nat}$,
critical spreading occurs at
$p = p_c > p_{c, bulk}$, where an
active steady state is not possible.  As the population of open sites
spreads, it leaves in its wake a region devoid of open sites,
with particle density $\phi_f$, slightly greater than
$\phi_{nat}$.  After some time,
this region cannot be reactivated.  (For $p > p_{c, nat}$,
the survival probability decays exponentially even when
$\phi = \phi_{nat}$.)  Thus
for $p_{c,nat} < p < p_c (\phi) $, the active region is confined to
an expanding ring;  it is a kind of ``chemical wave" which converts one
type of absorbing configuration to another as it passes.
An example of this kind of spreading is shown in Fig 8.
The mean density of open sites (averaged over 500 trials),
is plotted in Fig. 9 as a function of distance
from the origin, confirming that the active region forms an
expanding ring.  Critical spreading with $\phi > \phi_{nat}$ presents a
rather different picture. Here the critical
point lies {\em below} $p_{c,bulk}$, which means that
surviving trials consist of an expanding region
with a small but finite density of open sites.

For non-natural initial conditions, the exponents
violate the hyperscaling relation, Eq (\ref{ghyper}).  This is not
surprising, since the scaling argument assumes that surviving processes
evolve, for large $t$, into a bulk critical system.\cite{mendes,torre}
This assumption, as we have seen, is not valid when $p_c \neq
p_{c, bulk}$.  To describe this novel situation, the scaling
analysis can be modified as follows.
We assume, as usual,  that spreading
may be described in terms of a pair of scaling functions,
defined {\em via} \cite{torre}

\begin{equation}
\label{sc1}
\rho(x,t) \sim t^{\eta - dz/2} G(x^{2}/t^{z}, \Delta t^{1/\nu_{||}}),
\end{equation}
and
\begin{equation}
\label{gensp}
P(t) \sim t^{-\delta} \Phi(\Delta t^{1/\nu_{||}}),
\end{equation}
where $\Delta = p_c (\phi) - p$ is the distance from the
critical point.  (For $\Delta >0$ spreading may continue indefinitely.)
In Eq. (\ref{sc1}) $\rho(x,t)$ is the local order-parameter density,
averaged over all realizations; it is concentrated near the
origin at time zero.
The prefactors are constructed to yield power laws ---
Eqs. (\ref{survpwr}) - (\ref{sprpwr}) --- when $\Delta = 0$.
Existence of the limit $P_{\infty}$
implies that $\Phi (x) \sim x^{\beta'} $ as $x \rightarrow \infty$,
with $\beta' = \delta \nu_{||}$.

Consider spreading with $\phi < \phi_{nat}$ and
$p_c (\phi) > p_{c, bulk}$, in which the active region is an
expanding ring.
Simulations indicate that for $\Delta > 0$, the active region expands
at a constant speed $v(\Delta)$, and that the maximum open-site
density in surviving trials attains a steady value $\rho_o (\Delta)$
for large $t$.
In the contact process and similar models,
$\rho({\bf x},t) \rightarrow P_{\infty} \overline{\rho}$ as
$t \rightarrow \infty$, for any fixed ${\bf x}$.  But here we expect
instead that $\rho({\bf x},t) \rightarrow P_{\infty}
\rho_o (\Delta) f(|{\bf x}|/vt)$,
where $f(u)$ attains its maximum value
(unity) at $u_0 \approx 1$, and $f \rightarrow 0$
as $u \rightarrow 0 $ or $u \rightarrow \infty$.  We expect
$\rho_0 $ to have a power-law dependence on $\Delta$:
\begin{equation}
\label{betapp}
\rho_0 \sim \Delta ^{\beta''}.
\end{equation}

\noindent Consider the limit
$t \rightarrow \infty$, $|{\bf x}| \rightarrow \infty$,
with $|{\bf x}|/vt \approx u_0$. For $\Delta $ small but positive,
$\rho({\bf x},t) \sim \Delta^{\beta'} \rho_0 $, so
that $G(\infty,y) \sim y^{\beta' +\beta''} $ for large $y$.
On the other hand, we must have $G(\infty,y) \sim y^{-\nu_{||}
(\eta - dz/2)} $
for $\lim_{|{\bf x}|, t \rightarrow \infty} \rho({\bf x},t) $
to exist.  Comparing
these asymptotic behaviors, we find a
hyperscaling relation for annular spreading
\begin{equation}
\label{hyperann}
(1 +\frac{\beta''}{\beta'})\delta + \eta = \frac {dz}{2}.
\end{equation}
Inserting the exponent values shown in Table 1 yields $\beta'' =
1.80(35)$, for
$\phi = 0.2$.  This is consistent with
$\beta'' = 1.68(6)$, obtained directly from simulations
at $p$-values between $p_{c, bulk} $ and
$p_c  (0.2) $.

As noted above, surviving trials at $p_c(\phi)$ for $\phi > \phi_{nat}$
develop a finite open-site density and so are compact objects.
The critical exponents for
spreading of compact colonies are expected to satisfy
the relation \cite{hscdp}:

\begin{equation}
\label{hyperdisc}
\delta + \eta = \frac {dz}{2},
\end{equation}
\noindent which simply expresses the growth of population
in a region of positive density, given that the radius grows
$ \sim t^{z/2}$.
The exponents
for density $\phi =0.5$ are, however, inconsistent
with Eq. (\ref{hyperdisc}): $\delta + \eta - z = 0.27(13)$
(The deviation from the generalized DP hyperscaling relation,
Eq. (\ref{ghyper}), is considerably larger:
$ (1+\beta/\beta')\delta + \eta - z = 0.6(2)$.)
For RSA initial configurations ($\phi = 0.36413$)
the corresponding deviations are -0.3(1) and 0.2(1).
We can understand this discrepancy by noting that
the onset of compact growth, which requires that
the colony diameter $D >> \xi$, the
bulk correlation length, should only
occur at very long times, since $|p - p_{c, bulk} |/p_{c, bulk}$
is very small
(less than $10^{-3}$).  This is supported by the observation that
even at fairly late times ($t = 10^4$), the distribution of
open sites does not appear compact, but rather
consists of several disconnected
regions (see Fig. 10).  Presumably the spreading exponents
found for $\phi =0.5$ and 0.364 are not the
asymptotic ones; their determination will require studies of
considerably longer duration.

\section{Summary}

The first study of nonuniversal spreading in two
dimensions reveals a large variation of critical exponents
with the density $\phi$ of the environment into which the process grows.
The variation of the spreading exponents
is much more pronounced
than in one-dimensional models with multiple absorbing
configurations \cite{pcppre,mendes}.
In the SNR $\delta $ and $\beta'$ vary by
more than an order of magnitude, $\eta$ changes sign, and $z$, which
is essentially constant in the one-dimensional models, varies by more
than a factor of two. (Similarly, $\delta + \eta$ appears constant in
one dimension, but varies substantially here.)
Moreover, the critical point, $p_c$, depends
on $\phi$, whereas in one dimension it remains constant.
The shift in $p_c$ has important consequences for scaling since
the critical spreading process does not evolve into a bulk
critical state for $\phi \neq \phi_{nat}$.
Comparison of spreading in two
environments with the same density but different symmetries reveals
that the process is strongly influenced
by factors other than the density.

\acknowledgments

I thank Geoff Grinstein and Miguel Angel Mu\~{n}oz for
helpful discussions.
Simulations employed the facilities of
the University Computing Center, City University of New York.


\newpage
\noindent{\bf Figure Captions}
\vspace{1em}

\noindent Fig. 1.  Dependence of the half-life, $\tau$, on system size.
$\nu_{||}/\nu_{\perp} = 1.764$, as expected for DP in 2+1 dimensions.
Filled squares: $p= 0.3928$; $\Box$: $p= 0.3927$; $\times$: $p= 0.3926$;
$\diamond$: $p= 0.3925$.
\vspace{1em}

\noindent Fig. 2. Survival probability, mean number of open sites, and
mean-square spread for natural initial conditions, $p=0.392615$.
\vspace{1em}

\noindent Fig. 3.  Local slopes, $\delta_t$ (a), $\eta_t$ (b), and
$z_t$ (c), for natural initial conditions. $\diamond$: $p=0.39261$;
$\Box$: 0.392615; $\times$: 0.39262; +: 0.39263.
\vspace{1em}

\noindent Fig. 4. Two-point correlation function $h(r)$ in natural
absorbing configurations.
$\bullet$: (1,0) direction; $\times$: (1,1) direction; $\circ$:
(2,1) direction.
\vspace{1em}

\noindent Fig. 5.  Examples of initial absorbing configurations
investigated in this work.
Left: upper, density $\phi = 1/5$; mid, $\phi = 1/4$, anisotropic;
lower, $\phi = 0.363$,
(RSA); right: upper, $\phi = 0.328$, (natural); mid, $\phi = 0.25$,
isotropic; lower,
$\phi = 0.5$.
\vspace{1em}

\noindent Fig. 6.  Local slopes, $-\delta_t$ (a), $\eta_t$ (b),
and $z_t$ (c),
for $\phi = 0.2$.  $+$: $p=0.4106$;  $\Box$: $p=0.4104$;
$\diamond$: $p=0.4102$;
$\times$: $p=0.4100$;  $\circ$: $p=0.4096$.
\vspace{1em}

\noindent Fig. 7. Spreading exponents $\delta$, $\eta$, $z$, and
$\beta'$ vs initial density $\phi$.

\noindent Fig. 8. Spread of open sites, $\phi = 0.2$, $p=p_c$.  Light
gray: $t=200$; dark gray: $t=500$; black: $t=1000$.
\vspace{1em}

\noindent Fig. 9.  Mean density of open sites vs distance from seed, for
$\phi = 0.2$, $p=p_c$.  From left to right, $t= 200, 500$, and 1000.
\vspace{1em}

\noindent Fig. 10. A typical colony for $\phi = 0.5$, $p = p_c$,
$t=10^4$. The line represents 100 lattice spacings.
\vspace{1em}

\end{document}